%% file: mainprb.tex
\documentclass[aps,prb,reprint,twocolumn,showpacs,floatfix,superscriptaddress]{revtex4-2}

\usepackage{amssymb,amsmath,amstext}                
\usepackage{graphicx}                                               
\usepackage{epstopdf}                                               
\usepackage{color}                                                     
\usepackage{bm}                                                        
\usepackage{appendix}                                              
\usepackage[utf8]{inputenc}
\usepackage{latexsym}
\usepackage{xcolor}
\usepackage{braket}
\usepackage{ulem}
\usepackage{siunitx}
\usepackage{umoline}
\usepackage{epsfig}
\usepackage{graphics}
\usepackage{amssymb}
\usepackage{psfrag}
\usepackage{mathtools}
\usepackage[english]{babel}
\usepackage[T1]{fontenc}
\usepackage{lmodern}
\usepackage{booktabs}

\usepackage{comment}

\usepackage[colorlinks, linkcolor=blue, anchorcolor=blue, citecolor=blue]{hyperref}

\begin{document}

\title{Ground states of one-dimensional dipolar lattice bosons at unit filling}

\author{Mateusz \L\c{a}cki} 
\affiliation{Institute of Theoretical Physics, Jagiellonian University in Krakow, ul. Lojasiewicza 11,
	30-348 Krak\'ow, Poland}

\author{Henning Korbmacher}
\affiliation{Institut f\"ur Theoretische Physik, Leibniz Universit\"at Hannover, Germany}

\author{Gustavo A. Dom\'inguez-Castro}
\affiliation{Institut f\"ur Theoretische Physik, Leibniz Universit\"at Hannover, Germany}

\author{Jakub Zakrzewski}
\email{jakub.zakrzewski@uj.edu.pl}
\affiliation{Institute of Theoretical Physics, Jagiellonian University in Krakow, ul. Lojasiewicza 11,
	30-348 Krak\'ow, Poland}
\affiliation{Mark Kac Complex Systems Research Center,  Jagiellonian University in Krakow, \L{}ojasiewicza 11, 30-348 Krak\'ow, Poland}

\author{Luis Santos}
\affiliation{Institut f\"ur Theoretische Physik, Leibniz Universit\"at Hannover, Germany}

\date{\today}

\begin{abstract}
Recent experiments on ultracold dipoles in optical lattices open exciting possibilities for the quantum simulation of extended Hubbard models. When considered in one dimension, these models present at unit filling a particularly interesting ground-state physics, including a symmetry-protected topological phase known as Haldane insulator. We show that the tail of the dipolar interaction beyond nearest-neighbors, which may be tailored by means of the transversal confinement, does not only modify quantitatively the Haldane insulator regime and lead to density waves of larger periods, but results as well in unexpected insulating phases. These insulating phases may be topological or topologically trivial, and are characterized by peculiar correlations of the site occupations. These phases may be realized and observed in state-of-the-art experiments.
\end{abstract}

\maketitle




\section{Introduction}

Ultra cold quantum systems in optical lattices and tweezer arrays offer extraordinary possibilities for the quantum simulation of many-body lattice models~\cite{Lewenstein07,Bloch2008,Lewenstein12}. Whereas these systems are typically characterized 
by on-site interactions, with the exception of relatively weak super-exchange terms~\cite{Trotzky2008}, recent experiments on dipolar systems, including 
magnetic atoms~\cite{Chomaz2023}, polar molecules~\cite{Bohn2017}, and Rydberg gases~\cite{Browaeys2020}, open 
the possibility to simulate lattice models with significant inter-site interactions. Extended Hubbard models \cite{Dutta15} have been already realized using magnetic atoms~\cite{Baier2016, Su2023}, including the very recent observation of density-wave patterns~\cite{Su2023}.

 One-dimensional extended Bose-Hubbard models~(EBHMs) 
 are particularly interesting at unit filling, since they 
 present in addition to 
the superfluid (SF), Mott insulator~(MI) and density-wave (DW) phases, the so-called Haldane insulator (HI)~\cite{DallaTorre2006, Rossini2012, Ejima2014}, a symmetry protected topological phase, characterized by a string order, and a doubly-degenerate entanglement spectrum~\cite{Pollmann2010}. This phase is directly 
linked to the Haldane phase of spin-$1$ 
chains~\cite{Nijs1989,Kennedy1992}, since for strong-enough on-site interactions, EBHMs may be mapped to spin-$1$ models~\cite{DallaTorre2006}. Recently it was shown \cite{Fraxanet22} that topological properties can persist at the HI-DW phase transition.

Although major features of the physics of polar lattice gases may be understood using models with only nearest-neighbor interactions, 
for sufficiently large dipole moments, the tail of the dipolar interactions beyond nearest neighbors, which is typically assumed as decaying as $1/r^3$, with $r$ the inter-site distance, may become relevant~\cite{DallaTorre2006,Biedron18,Kraus20}. 
The dipolar tail leads to quantitative modifications of the HI parameter region, but it does not compromise its existence~\cite{DallaTorre2006}. In addition, it is expected that strong-enough interactions beyond nearest neighbors should lead to DWs with longer integer spatial periods. 


\begin{figure}[t!]
\includegraphics[width=\columnwidth]{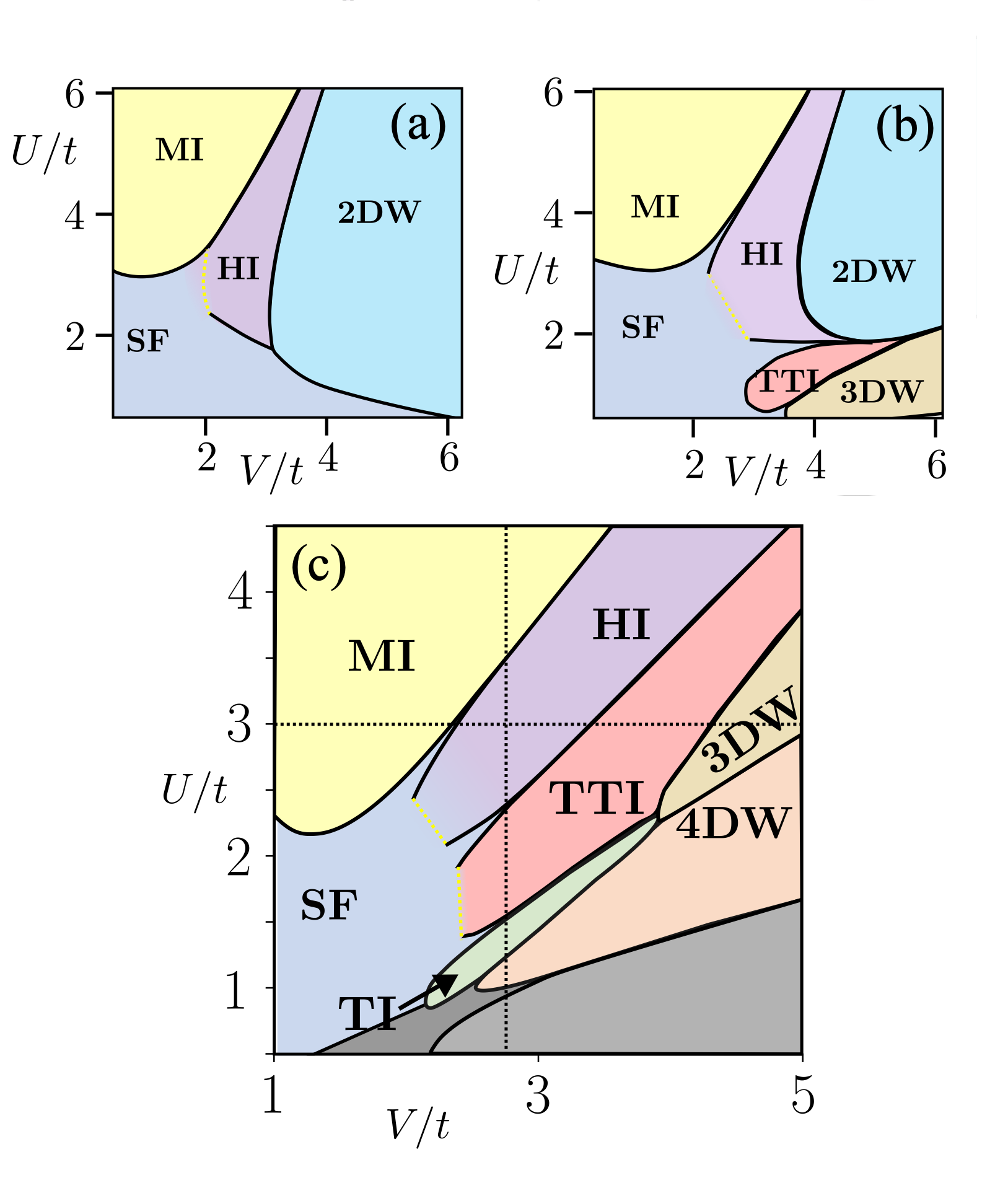}
	\caption{Ground-state phase diagrams as a function of the on-site interactions $U/t$ and nearest-neighbor interaction $V/t$, for (a) nearest-neighbor model, (b) $1/r^3$ dipolar tail, and (c) dipolar tail for $\beta_{eff}=1$ ($B\approx 0.297$). In panel (c), dotted black lines indicate the cuts evaluated in detail in Fig.~\ref{fig:2}, and the grey area is the region in which our iDMRG does not provide accurate results due to the unit cell and the maximal occupation per site considered. 
  In all panels, dotted yellow boundaries indicate approximate HI-SF and TTI-SF boundaries which could not be fully established in our iDMRG numerics. 
    }
	\label{fig:1}
\end{figure} 


In this work, we show that, surprisingly, the dipolar tail 
results as well in the appearance of unexpected insulating phases. For the usually considered $1/r^3$ dipolar tail, we 
find at low on-site interactions a topologically trivial spatially disordered insulator~(TTI). Moreover, by properly choosing the transversal confinement it is possible to taylor a milder  decay of the dipolar tail~\cite{Wall2013, Korbmacher2023,PhysRevA.107.063307}, which results not only in an enhanced TTI region, but also in an additional topological insulator~(TI) phase, characterized by a doubly-degenerate entanglement spectrum. This phase resembles the HI, but it cannot be trivially mapped to an integer spin chain. We show that the TTI and TI phases present peculiar correlations of the site occupations that we characterize using generalized string correlations, which may be measured using quantum gas microscope techniques~\cite{Gross2021}.
We characterize the corresponding phase transitions. In particular, the TTI~(TI) phases undergo, for large-enough dipolar strength, a second-order transition 
belonging to the 3-state~(4-state) Potts universality class into a density wave with a period of 3~(4) sites. These phases may be realized and observed in state-of-the-art experiments.


The paper is organized as follows. In Sec.~\ref{sec:Model} we introduce the extended Hubbard model that characterizes the polar lattice gas. Section~\ref{sec:NN} reviews the phase diagram for the nearest-neighbor model.
Section~\ref{sec:PolarLattice} discusses the phase diagram for polar lattice gases, and in particular the properties of the TTI and TI phases, as well as the involved phase transitions. Finally, we conclude in Sec.~\ref{sec:Conclusions}.



\section{Model}
\label{sec:Model}

We consider dipolar bosons of mass $m$ in a 1D optical lattice $V_0\sin^2(\pi z/a)$, where $V_0$ is the lattice depth and $a$ is the lattice spacing, of length $L$, confined transversally by an isotropic harmonic potential $\frac{1}{2}m\omega_\perp^2(x^2+y^2)$, where $\omega_\perp$ is the trap frequency. The system is  described by an EBHM

\begin{eqnarray}
\hat{H} &=& 
-t\sum_i \left(\hat{b}^\dagger_{i+1}\hat{b}_i+\mathrm{H.c.}\right) 
\nonumber \\
&+& \frac{U}{2}\sum_i \hat{n}_i\left(\hat{n}_i\!-1\right)+
\sum_i\sum_{j>0} V_j\hat{n}_i\hat{n}_{i+j},
\end{eqnarray}
where $t$ is the (nearest-neighbor) hopping amplitude, $\hat{b}^\dagger_i$ ($\hat{b}_i$) is the creation (annihilation) operator at site $i$, $\hat{n}_i=\hat{b}^\dagger_i\hat{b}_i$ the corresponding number operator, and $U$ characterizes the on-site interactions. 

The dipole-induced inter-site interactions are given by $V_j=VG_j(B)$~\cite{Korbmacher2023}, 
where $V$ is the interaction strength 
between nearest-neighboring sites, and the decay of the dipolar tail is given by 
\begin{equation}
    G_j(B)=f(\sqrt{B}j)/f(\sqrt{B}), 
\end{equation}
with
\begin{equation}
    f(\xi)=2\xi-\sqrt{2\pi}(1+\xi^2)e^{\xi^2/2}\mathrm{erfc}(\xi/\sqrt{2}),
\end{equation}
where 
\begin{equation}B=\frac{\pi^2 \zeta}{2} (1-\frac{\zeta}{2\sqrt{s}})^{-1},
\end{equation}
with $\zeta=\hbar\omega_\perp/E_R$, $s=V_0/E_R$, and  $E_R=\frac{\pi^2\hbar^2}{2ma^2}$ the recoil energy. 
Although the dipolar tail eventually decays for large $j$ as $1/j^3$, at short distances it may significantly depart from that dependence due to the transversal confinement. This departure is uniquely characterized by the exponent $\beta_\text{eff}$, with $V_2/V=1/2^{\beta_\text{eff}}$ ~\cite{Korbmacher2023,PhysRevA.107.063307}. 
 The case $\beta_\text{eff}=3$ corresponds to the usually assumed $V_j=V/j^3$ at all distances (corresponding to large $B\approx 100$, dependent on the value of $s$). With lower $\beta_\text{eff}$, $V_j$ may be significantly enhanced, modifying the ground-state properties, as discussed below, in particular for  $\beta_\text{eff}=1$ (corresponding to $B\approx 0.297$).
 
 In the following, we evaluate the ground state at unit filling  using the infinite density-matrix renormalization group~(iDMRG) technique~\cite{Mcculloch08}, with bond dimensions up to $\chi=1200$,  
 and a unit cell of $12$ sites, truncating the inter-site interactions $V_j$ at $20$ sites. Since we are interested in regimes of low $U/t$ values, we consider in our numerical simulations up to $5$ particles per site~(larger occupations do not modify our conclusions). We assume that the lattice is deep-enough, such that we can neglect the effect of density-assisted hopping~\cite{Sowinski2012,Dutta15,Kraus22} as well as the effects due to higher bands \cite{Lacki13,Hughes23}.



\section{NEAREST-NEIGHBOR MODEL} 
\label{sec:NN}

For the case of nearest-neighbor interactions, $V_j=V\delta_{j,1}$~(Fig.~\ref{fig:1}(a)), the ground-state phase diagram splits into four 
phases~\cite{Rossini2012}~(we do not 
consider the regime of low $U/t$ and large $V/t$, characterized by phase separation~\cite{Batrouni2014, Kottmann2021}): a SF phase at low $U/t$ and $V/t$, characterized by power-law decaying single-particle correlations
$C_\text{SP}(i,j)=\langle\hat{b}^\dagger_i\hat{b}_{j}\rangle$, and three insulating phases~(with exponentially decaying $C_\text{SP}$): MI, DW with a period of two sites~(2DW), and HI. These insulators relate, respectively, with the large-$D$, N\'eel, and Haldane phases of spin-$1$ chains with single-ion anisotropy~\cite{Chen2003}. This is because at large-enough $U/t$ only site occupations $0$, $1$, and $2$ play a significant role
~\cite{DallaTorre2006}. 

The MI displays finite parity order
\begin{equation}
    \mathcal{O}_P=\lim_{|i-j|\rightarrow\infty}\langle e^{i\pi\sum_{i<k<j}\delta\hat{n}_k}\rangle.
\end{equation}
with $\delta\hat{n}_i=\hat{n}_i-1$, and vanishing string order~\cite{Nijs1989}
\begin{equation}
    \mathcal{O}_S=\lim_{|i-j|\rightarrow\infty}\langle\delta\hat{n}_ie^{i\pi\sum_{i<k<j}\delta\hat{n}_k}\delta\hat{n}_j\rangle.
\end{equation}
The 2DW phase, occurring at sufficiently large $V/t$, shows a staggered density-density correlation revealed by a peak at momentum $k=\pi$ 
in its structure factor
\begin{equation}
    S(k)=\frac{1}{L^2}\sum_{i,j}e^{ik(i-j)}\langle\hat{n}_i\hat{n}_j\rangle,
\end{equation}
with $L$ the number of sites. Finally, the HI phase, occurring at intermediate $U/t$ ad $V/t$ values, presents a finite ${\cal O}_S$ and vanishing ${\cal O}_P$. 
The HI is further characterized 
by its entanglement spectrum $\{\lambda_i\}$, i.e. the eigenvalues of the reduced density matrix $\rho_A=\mathrm{Tr}_B\rho$, obtained from the overall density matrix $\rho$ when partitioning the system into two subsystems $A$ and $B$. The HI is a symmetry-protected topological phase with a double-degenerate entanglement spectrum~\cite{Pollmann2010}, i.e. $P_\Lambda=0$, with  \begin{equation}
P_\Lambda=\sqrt{\sum_{i} \left (\lambda_{2i}-\lambda_{2i+1} \right )^2},
\end{equation}
where the sum is over all eigenvalues of $\rho_A$.



\begin{figure}[t!]%
\includegraphics[width=\columnwidth]{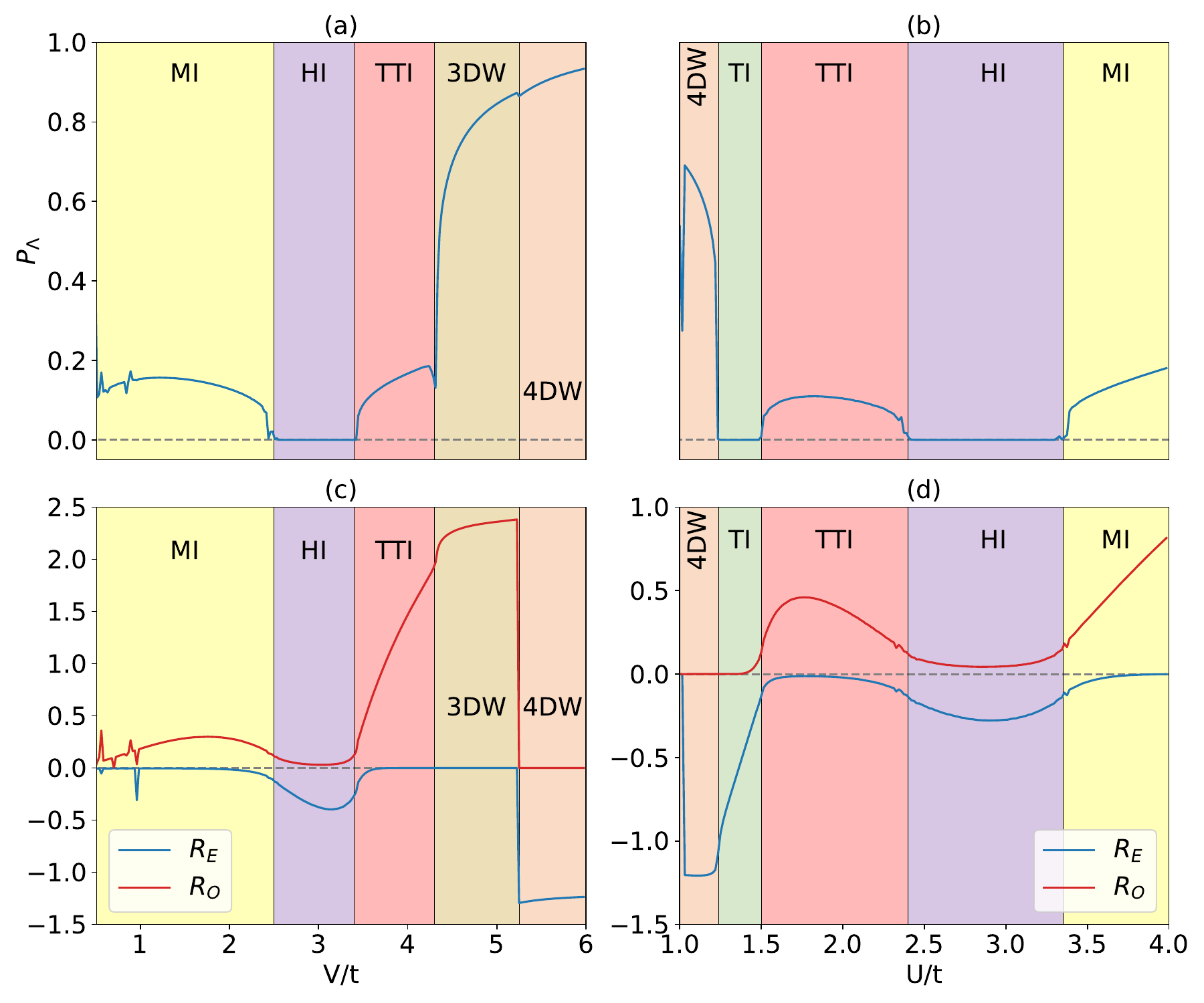}
\caption{(a) Degeneracy of the entanglement spectrum, $P_\Lambda$, as a function of $V/t$ for $\beta_{\textrm{eff}}=1$ ($B\approx 0.297$) and $U/t=3$~(dotted horizontal line in Fig. ~\ref{fig:1}~(c). 
(b) The same $P_\Lambda$ shown as a function of $U/t$ for $V/t=2.75$~(dotted vertical line in Fig.~\ref{fig:1}(c)). (c) String correlations $R_E$ and $R_O$~(see text) for the parameter range of panel (a). (d) Same as (c) but for parameters of panel (b).}
\label{fig:2}
\end{figure} 



\section{POLAR LATTICE GASES}
\label{sec:PolarLattice}

The dipolar tail modifies quantitatively the extension of the HI parameter region~(panels (b) and (c) of Fig.~\ref{fig:1}), but does not compromise its existence~\cite{DallaTorre2006}. At large $V/t$, additional DW phases appear, with period $\ell=3$~(3DW), $4$~(4DW), and so on, characterized by peaks at $k=2\pi/\ell$ in the structure factor $S(k)$. The $\ell$DW and $(\ell+1)$DW phases are separated from each other by a first-order phase transition at $(U/V)_\ell = 2 
\sum_{s>0} \left [nG_{s\ell}-(\ell+1) G_{s(\ell+1)}\right ]$. 
Interestingly, for low $U/t$ values, the dipolar tail results in additional phases 
which we term TTI~(topologically-trivial insulator) and TI~(topological insulator), 
which to the best of our knowledge have remained up to now unnoticed. 
The TTI phase is apparent even for the case of an unmodified $1/j^3$ dipolar tail~(Fig.~\ref{fig:1}~(b)). Decreasing $\beta_\textrm{eff}$ enhances the 
TTI region and reveals the TI phase, see Fig.~\ref{fig:1}~(c). 
We discuss below the nature of these unexpected phases.



\subsection{Topologically Trivial Insulator phase} 

\begin{figure*}%
\includegraphics[width=2\columnwidth]{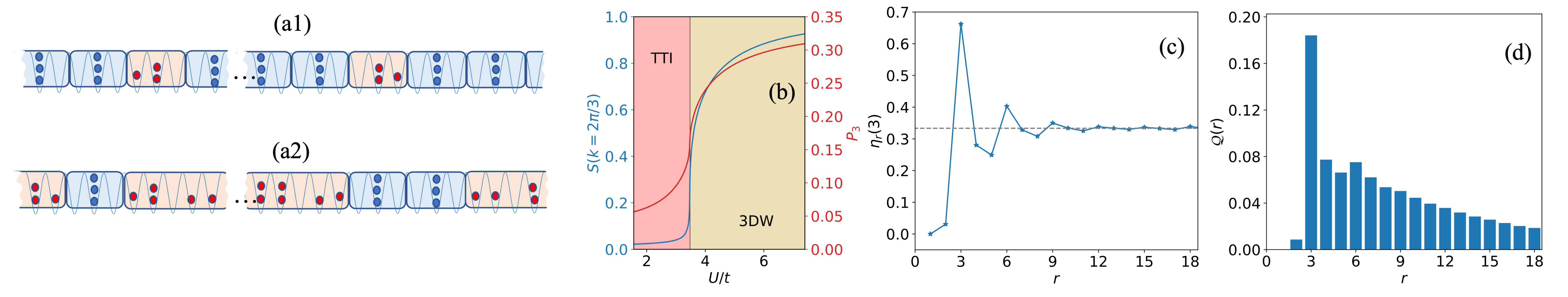}
\caption{(a1) Sketch of the number distribution deep within the 3DW phase: $(030)$ blocks~(blue slabs) with occasional interspersed defects $(021)$ or $(120)$~(red slabs)induced by hopping. (a2) In the TTI, regions with holons, singlons, and doublons~(red slabs) are interspersed with a low filling of $(030)$ blocks~(blue slabs), which retain a tendency to form pairs.
(b) Structure factor $S(k=2\pi/3)$ and triplon population $P_3$ at the TTI-3DW interface for $U/V=0.74$. 
(c) Triplon correlation $\eta_r(3)$ and (d) 
probability $\mathcal{Q}_r$ to find nearest triplons at a distance $r$~(see text), for $U/t=2.8$ and $V/t=3.8$, well within the TTI phase. In all cases we consider $\beta_{\mathrm{eff}}=1$.}
\label{fig:3}
\end{figure*}

The TTI phase is an insulator (with exponentially decaying $C_{SP}$) sandwiched between the 3DW and the HI phase. In contrast to the HI,  
the entanglement spectrum is not doubly degenerate, i.e.
$P_\Lambda\neq 0$~(see panels (a) and ~(b) of Fig.~\ref{fig:2}). 
Also, the TTI presents no structure factor peak at $k\neq 0$. In particular, the peak at $k=2\pi/3$, characterizing the 3DW, vanishes at the TTI-3DW transition~(Fig.~\ref{fig:3}(a)). 
Despite of the lack of spatial ordering, the TTI displays intriguing 
correlations of the site occupations, which we characterize by means of the generalized string correlations~\cite{footnote-R} 
\begin{equation}
   R(q)=\lim_{j\to\infty}\frac{\langle\hat{P}_0(q)e^{i\pi\sum_{0<k<j}\sum_s\hat{P}_k(2s)}\hat{P}_j(q)\rangle}{\langle \hat{P}_0(q)\hat{P}_j(q)\rangle},
\end{equation}
where $\hat{P}_i(q)=\prod_{k\neq s}(\hat{n}-k)/\prod_{k'\neq q}(q-k')$ 
project into states with $q$ particles at site $i$. 
The correlation $R(q)$ measures 
the parity of the number of evenly occupied sites between two sites occupied by $q$ particles. We may characterize the insulating phases by monitoring $R_E=\sum_s R(2s)$ and $R_O=\sum_s R(2s+1)$, as illustrated in Fig.~\ref{fig:2}(c) and Fig.~\ref{fig:2}(d). In particular, the HI phase, with its diluted antiferromagnetic order of alternated empty sites~(holons) and doubly-occupied ones~(doublons) within 
a set of otherwise singly-occupied sites~(singlons), is characterized by $R_O\simeq 0$ and $R_E<0$.


 
Deep in the 3DW phase, i.e. at large $V/t$ for $(U/V)_2<(U/V)<(U/V)_3$, the ground state is approximately a crystal of triply-occupied sites~(triplons) $|\dots 030030030 \dots\rangle$, with maximal triplon population $P_3\equiv\sum_i \langle \hat P_i(3) \rangle = L/3$. Due to hopping, $(030)$ blocks 
occassionally decay into defects $(120)$ or $(021)$ as depicted in Fig.~\ref{fig:3}(a1). 
The 3DW is hence characterized by $R_O>0$, since between two triplons there is always an even number of doublons plus holons. Moreover, $R_E=0$, 
due to the arbitrary parity of the number of evenly-occupied sites between two holons or two doublons.

When $V$ decreases the number of defects in the 3DW increases. 
At a critical $V$, which coincides with half triplon filling,  $P_3\simeq L/6$ 
 -- Fig.~\ref{fig:3}(b) -- 
the system enters the TTI phase.  
Defects $(021)$ an $(120)$ unravel due to hopping, and the parity 
order associated to pairs $(12)$ or $(21)$, $\mathcal{O}_P(1,2)=\langle (-1)^{\sum_{i<l<j}(P_1(l)+P_2(l))}\rangle$ which is finite inside the 3DW, vanishes at the 3DW-TTI transition. 
However, most hops only affect holons, singlons and doublons located between $(030)$ blocks~(orange region in Fig.~\ref{fig:3}~(a2)), which do not modify $R_E$ or $R_O$. As a result,  correlations $R_E\simeq 0$ and $R_O>0$ are preserved in the TTI phase, compare panels~(c) and~(d) in Fig.~\ref{fig:2}.

Moreover, although the long-range 3DW order is lost, short-range triplon correlations
$\eta_r(3)=\frac{1}{3}\langle\hat{P}_i(3)\hat{P}_{i+r}(3)\rangle/ \langle\hat{P}_i(3)\rangle^2$
are still significant, see Fig.~\ref{fig:3}(c). For comparison, note that $\eta_{3n}(3)=1$ and $\eta_{3n\pm 1}(3)=0$ deeply inside the 3DW phase. Further understanding is achieved by monitoring the probability that nearest-neighboring triplons are $r$ sites apart, 
$\mathcal{Q}_r= \bar{\mathcal{Q}}_r/\sum_r  \bar{\mathcal{Q}}_r$, with 
\begin{equation}
 \bar{\mathcal{Q}}_{r}=\langle\hat{P}_j(3)[\Pi_{j<k<j+r}(1-\hat{P}_k(3))]\hat{P}_{j+r}(3)\rangle.   
\end{equation}
Deep in the 3DW phase, $\mathcal{Q}_{r}$ presents obviously a single peak at $r=3$. When $V/t$ decreases, the formation of defects leads to finite $\mathcal{Q}_{6, 9, \dots}$. Upon a further decrease of $V/t$, triplon destruction leads to an approximately random 
separation between nearest triplons in the TTI phase, resulting in the exponential tail in the $\mathcal{Q}_r$ distribution observed in Fig.~\ref{fig:3}(d)). At short distances the triplon distribution is however not random. Triplons strongly block each other up to two sites apart. Moreover, triplons in the TTI phase still keep a marked tendency to form pairs placed three sites apart, as also indicated by the strong correlation $\eta_3(3)$ observed in Fig.~\ref{fig:3}(c). All these correlations vanish when the TTI transitions into the SF.



\subsection {Topological Insulator phase}

Remarkably, whereas reducing $V/t$ induces the melting of the 3DW  into the TTI, melting of the 4DW leads to a qualitatively different~(TI) phase. As for the TTI, the TI phase is 
an insulator with exponentially-decaying $C_\text{SP}$ and has no 
spatial order. In particular, the peak of $S(k)$ 
at $k=\pi/2$ characterizing the 4DW phase vanishes when entering the TI. Deep in the 4DW phase, the state $|\dots 004000400 \dots\rangle$ fulfills $R_E<0$ and $R_O=0$. Hopping results in defects $(31)$ and $(13)$, which do not affect these correlations. At the 3DW-TI phase transition the pairing order $\mathcal{O}_P(1,3)=\langle (-1)^{\sum_{i<l<j}(P_1(l)+P_3(l))}\rangle$ vanishes, 
however,  the $R_E$ and $R_O$ correlations are maintained, as for the TTI phase
with $R_E<0$ and $R_O\simeq 0$, as for the HI. Remarkably, again as for the HI, the TI displays a doubly-degenerate entanglement spectrum, see Fig~\ref{fig:2}~(b). However, in contrast to the HI where only holons, singlons and doublons played a relevant role, the TI presents a large probability to display triplons and four-fold occupied sites. As a result, the string-order 
$\mathcal O_S$ vanishes in the TI phase.



\subsection{Phase transitions}

\begin{figure}[t!]%
\includegraphics[width=0.8\columnwidth]{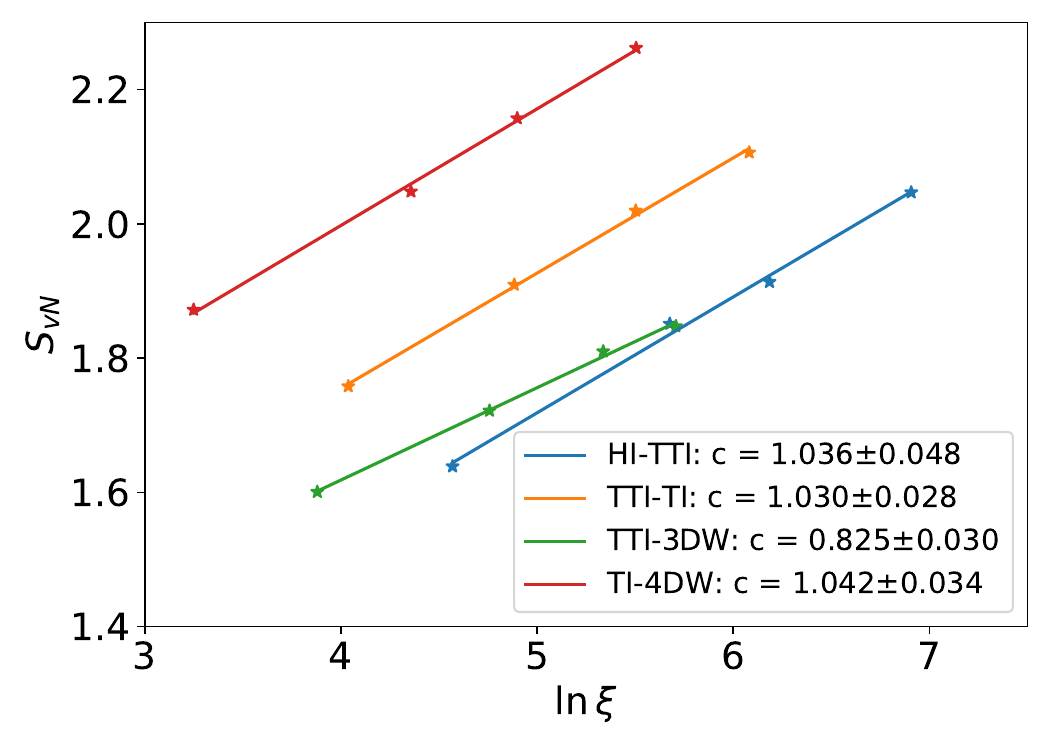}
\caption{Entanglement entropy scaling with the correlation length obtained from iDMRG calculations for the HI-TTI and TTI-3DW transitions~(obtained, respectively, for $U/t=3$, $V/t \approx 3.434$ and $U/t\approx 3.479$, $V/t\approx 4.688$), and for the TTI-TI and TI-4DW transitions~(obtained, respectively, for $U/t=3$, $V/t \approx 4.331$ and $U/t \approx 1.319$, $V/t\approx 2.481$). The numerical value extracted is in good agreement with 
$c=1$ for HI-TTI, TTI-TI and TI-4DW, and $c=4/5$ for TTI-3DW. In all cases we consider $\beta_{\mathrm{eff}}=1$.}
\label{fig:4}
\end{figure}

Whereas the direct transitions occurring at large-enough $U/t$ and $V/t$ between MI and 2DW and between the $\ell$DW phases are first-order, those between the insulating phases occurring at low $U/t$ and $V/t$ are second-order. These transitions are characterized by the behavior of the von Neumann entanglement entropy $S_{vN}$ at criticality, which we obtain from our iDMRG calculations, and which up to a constant acquires the form~\cite{Calabrese2008}:
\begin{equation}
S_{vN}(\chi)=\frac{c}{6}\log\xi_\chi,
\end{equation}
where $\chi$ is the bond dimension, $\xi_\chi$ is the $\chi$-dependent 
correlation length, and $c$ is the central charge.  As it is well known, the MI-HI transition is in the Luttinger liquid universality class with $c=1$~\cite{Berg2008, Ejima2014}. Our results show that the HI-TTI, and TTI-TI transitions are characterized as well by $c=1$~(see Fig.~\ref{fig:4}). As it is also known~\cite{Berg2008,Ejima2014}, we recover that the HI-2DW transition belongs to the Ising universality class with $c=1/2$. In contrast, the TTI-3DW transition is well described by the minimal conformal field theory with $c=4/5$, corresponding to the 3-state Potts universality class, whereas the TI-4DW transition 
is characterized by $c=1$, corresponding to the 4-state Potts universality class~\cite{Dotsenko2020}, see Fig.~\ref{fig:4}.



\section{Conclusions}
\label{sec:Conclusions}
For one-dimensional polar lattice gases at unit filling, the long-range tail of the dipolar interaction does not only result, as expected, in a quantitative modication of the Haldane-insulator parameter region, and in the existence of density waves with larger integer spatial periods, but also in the unexpected appearance of additional insulating phases. These spatially disordered phases present peculiar correlations of site occupations, and may be either topological~(TI) or topologically trivial~(TTI). 
For a $1/r^3$ dipolar tail, the TTI is found for $2\lesssim V/t\lesssim 5$, readily accessible in on-going erbium experiments~\cite{Su2023}. The observation of the TI demands softening the transversal confinement 
to modify the decay of the dipolar tail. Since this softening reduces $V/t$, the observation of the TI phase may demand the use of Dy in UV lattices, or the use of polar molecules, which should be both available in the near future.


\begin{acknowledgments}
M.\L. and J.Z. thank R. Kraus and G. Morigi for interesting discussions.
We acknowledge support of National Science Centre (Poland) via Opus grant 2019/35/B/ST2/00838 (M.\L{}.). This research was also funded by National Science Centre (Poland) under the OPUS call within the WEAVE programme 2021/43/I/ST3/01142 (J.Z.)
as well as by the
Deutsche Forschungsgemeinschaft (DFG, German Research Foundation) -- Project-ID 274200144 -- SFB 1227 DQ-mat within the project A04, FOR2247, and under Germany's Excellence Strategy -- EXC-2123 Quantum-Frontiers -- 390837967. 
We gratefully acknowledge Poland's high-performance Infrastructure PLGrid (HPC Centers: ACK Cyfronet AGH, PCSS, CI TASK, WCSS) for providing computer facilities and support within computational grant no. PLG/2022/015951. For the purpose of Open Access, the
authors applied a CC-BY public copyright licence to
any Author Accepted Manuscript (AAM) version arising from this submission.

\end{acknowledgments}

\input{mainprb.bbl}



\end{document}

%% file: mainprb.bbl
%

%% file: mainprb.bbl
\begin{thebibliography}{35}%
\makeatletter
\providecommand \@ifxundefined [1]{%
 \@ifx{#1\undefined}
}%
\providecommand \@ifnum [1]{%
 \ifnum #1\expandafter \@firstoftwo
 \else \expandafter \@secondoftwo
 \fi
}%
\providecommand \@ifx [1]{%
 \ifx #1\expandafter \@firstoftwo
 \else \expandafter \@secondoftwo
 \fi
}%
\providecommand \natexlab [1]{#1}%
\providecommand \enquote  [1]{``#1''}%
\providecommand \bibnamefont  [1]{#1}%
\providecommand \bibfnamefont [1]{#1}%
\providecommand \citenamefont [1]{#1}%
\providecommand \href@noop [0]{\@secondoftwo}%
\providecommand \href [0]{\begingroup \@sanitize@url \@href}%
\providecommand \@href[1]{\@@startlink{#1}\@@href}%
\providecommand \@@href[1]{\endgroup#1\@@endlink}%
\providecommand \@sanitize@url [0]{\catcode `\\12\catcode `\$12\catcode
  `\&12\catcode `\#12\catcode `\^12\catcode `\_12\catcode `\%12\relax}%
\providecommand \@@startlink[1]{}%
\providecommand \@@endlink[0]{}%
\providecommand \url  [0]{\begingroup\@sanitize@url \@url }%
\providecommand \@url [1]{\endgroup\@href {#1}{\urlprefix }}%
\providecommand \urlprefix  [0]{URL }%
\providecommand \Eprint [0]{\href }%
\providecommand \doibase [0]{https://doi.org/}%
\providecommand \selectlanguage [0]{\@gobble}%
\providecommand \bibinfo  [0]{\@secondoftwo}%
\providecommand \bibfield  [0]{\@secondoftwo}%
\providecommand \translation [1]{[#1]}%
\providecommand \BibitemOpen [0]{}%
\providecommand \bibitemStop [0]{}%
\providecommand \bibitemNoStop [0]{.\EOS\space}%
\providecommand \EOS [0]{\spacefactor3000\relax}%
\providecommand \BibitemShut  [1]{\csname bibitem#1\endcsname}%
\let\auto@bib@innerbib\@empty
\bibitem [{\citenamefont {Lewenstein}\ \emph {et~al.}(2007)\citenamefont
  {Lewenstein}, \citenamefont {Sanpera}, \citenamefont {Ahufinger},
  \citenamefont {Damski}, \citenamefont {Sen(De)},\ and\ \citenamefont
  {Sen}}]{Lewenstein07}%
  \BibitemOpen
  \bibfield  {author} {\bibinfo {author} {\bibfnamefont {M.}~\bibnamefont
  {Lewenstein}}, \bibinfo {author} {\bibfnamefont {A.}~\bibnamefont {Sanpera}},
  \bibinfo {author} {\bibfnamefont {V.}~\bibnamefont {Ahufinger}}, \bibinfo
  {author} {\bibfnamefont {B.}~\bibnamefont {Damski}}, \bibinfo {author}
  {\bibfnamefont {A.}~\bibnamefont {Sen(De)}},\ and\ \bibinfo {author}
  {\bibfnamefont {U.}~\bibnamefont {Sen}},\ }\bibfield  {title} {\bibinfo
  {title} {Ultracold atomic gases in optical lattices: mimicking condensed
  matter physics and beyond},\ }\href
  {https://doi.org/10.1080/00018730701223200} {\bibfield  {journal} {\bibinfo
  {journal} {Advances in Physics}\ }\textbf {\bibinfo {volume} {56}},\ \bibinfo
  {pages} {243} (\bibinfo {year} {2007})},\ \Eprint
  {https://arxiv.org/abs/https://doi.org/10.1080/00018730701223200}
  {https://doi.org/10.1080/00018730701223200} \BibitemShut {NoStop}%
\bibitem [{\citenamefont {Bloch}\ \emph {et~al.}(2008)\citenamefont {Bloch},
  \citenamefont {Dalibard},\ and\ \citenamefont {Zwerger}}]{Bloch2008}%
  \BibitemOpen
  \bibfield  {author} {\bibinfo {author} {\bibfnamefont {I.}~\bibnamefont
  {Bloch}}, \bibinfo {author} {\bibfnamefont {J.}~\bibnamefont {Dalibard}},\
  and\ \bibinfo {author} {\bibfnamefont {W.}~\bibnamefont {Zwerger}},\
  }\bibfield  {title} {\bibinfo {title} {Many-body physics with ultracold
  gases},\ }\href {https://doi.org/10.1103/RevModPhys.80.885} {\bibfield
  {journal} {\bibinfo  {journal} {Rev. Mod. Phys.}\ }\textbf {\bibinfo {volume}
  {80}},\ \bibinfo {pages} {885} (\bibinfo {year} {2008})}\BibitemShut
  {NoStop}%
\bibitem [{\citenamefont {Lewenstein}\ \emph {et~al.}(2012)\citenamefont
  {Lewenstein}, \citenamefont {Sanpera},\ and\ \citenamefont
  {Ahufinger}}]{Lewenstein12}%
  \BibitemOpen
  \bibfield  {author} {\bibinfo {author} {\bibfnamefont {M.}~\bibnamefont
  {Lewenstein}}, \bibinfo {author} {\bibfnamefont {A.}~\bibnamefont
  {Sanpera}},\ and\ \bibinfo {author} {\bibfnamefont {V.}~\bibnamefont
  {Ahufinger}},\ }\href@noop {} {\emph {\bibinfo {title} {Ultracold Atoms in
  Optical Lattices: Simulating quantum many-body systems}}}\ (\bibinfo
  {publisher} {Oxford University Press},\ \bibinfo {year} {Oxford,
  2012})\BibitemShut {NoStop}%
\bibitem [{\citenamefont {Trotzky}\ \emph {et~al.}(2008)\citenamefont
  {Trotzky}, \citenamefont {Cheinet}, \citenamefont {F\"olling}, \citenamefont
  {Feld}, \citenamefont {Schnorrberger}, \citenamefont {Rey}, \citenamefont
  {Polkovnikov}, \citenamefont {Demler}, \citenamefont {Lukin},\ and\
  \citenamefont {Bloch}}]{Trotzky2008}%
  \BibitemOpen
  \bibfield  {author} {\bibinfo {author} {\bibfnamefont {S.}~\bibnamefont
  {Trotzky}}, \bibinfo {author} {\bibfnamefont {P.}~\bibnamefont {Cheinet}},
  \bibinfo {author} {\bibfnamefont {S.}~\bibnamefont {F\"olling}}, \bibinfo
  {author} {\bibfnamefont {M.}~\bibnamefont {Feld}}, \bibinfo {author}
  {\bibfnamefont {U.}~\bibnamefont {Schnorrberger}}, \bibinfo {author}
  {\bibfnamefont {A.~M.}\ \bibnamefont {Rey}}, \bibinfo {author} {\bibfnamefont
  {A.}~\bibnamefont {Polkovnikov}}, \bibinfo {author} {\bibfnamefont {E.~A.}\
  \bibnamefont {Demler}}, \bibinfo {author} {\bibfnamefont {M.~D.}\
  \bibnamefont {Lukin}},\ and\ \bibinfo {author} {\bibfnamefont
  {I.}~\bibnamefont {Bloch}},\ }\bibfield  {title} {\bibinfo {title}
  {Time-resolved observation and control of superexchange interactions with
  ultracold atoms in optical lattices},\ }\href
  {https://doi.org/10.1126/science.1150841} {\bibfield  {journal} {\bibinfo
  {journal} {Science}\ }\textbf {\bibinfo {volume} {319}},\ \bibinfo {pages}
  {295} (\bibinfo {year} {2008})},\ \Eprint
  {https://arxiv.org/abs/https://www.science.org/doi/pdf/10.1126/science.1150841}
  {https://www.science.org/doi/pdf/10.1126/science.1150841} \BibitemShut
  {NoStop}%
\bibitem [{\citenamefont {Chomaz}\ \emph {et~al.}(2022)\citenamefont {Chomaz},
  \citenamefont {Ferrier-Barbut}, \citenamefont {Ferlaino}, \citenamefont
  {Laburthe-Tolra}, \citenamefont {Lev},\ and\ \citenamefont
  {Pfau}}]{Chomaz2023}%
  \BibitemOpen
  \bibfield  {author} {\bibinfo {author} {\bibfnamefont {L.}~\bibnamefont
  {Chomaz}}, \bibinfo {author} {\bibfnamefont {I.}~\bibnamefont
  {Ferrier-Barbut}}, \bibinfo {author} {\bibfnamefont {F.}~\bibnamefont
  {Ferlaino}}, \bibinfo {author} {\bibfnamefont {B.}~\bibnamefont
  {Laburthe-Tolra}}, \bibinfo {author} {\bibfnamefont {B.~L.}\ \bibnamefont
  {Lev}},\ and\ \bibinfo {author} {\bibfnamefont {T.}~\bibnamefont {Pfau}},\
  }\bibfield  {title} {\bibinfo {title} {Dipolar physics: a review of
  experiments with magnetic quantum gases},\ }\href
  {https://doi.org/10.1088/1361-6633/aca814} {\bibfield  {journal} {\bibinfo
  {journal} {Reports on Progress in Physics}\ }\textbf {\bibinfo {volume}
  {86}},\ \bibinfo {pages} {026401} (\bibinfo {year} {2022})}\BibitemShut
  {NoStop}%
\bibitem [{\citenamefont {Bohn}\ \emph {et~al.}(2017)\citenamefont {Bohn},
  \citenamefont {Rey},\ and\ \citenamefont {Ye}}]{Bohn2017}%
  \BibitemOpen
  \bibfield  {author} {\bibinfo {author} {\bibfnamefont {J.~L.}\ \bibnamefont
  {Bohn}}, \bibinfo {author} {\bibfnamefont {A.~M.}\ \bibnamefont {Rey}},\ and\
  \bibinfo {author} {\bibfnamefont {J.}~\bibnamefont {Ye}},\ }\bibfield
  {title} {\bibinfo {title} {Cold molecules: Progress in quantum engineering of
  chemistry and quantum matter},\ }\href
  {https://doi.org/10.1126/science.aam6299} {\bibfield  {journal} {\bibinfo
  {journal} {Science}\ }\textbf {\bibinfo {volume} {357}},\ \bibinfo {pages}
  {1002} (\bibinfo {year} {2017})}\BibitemShut {NoStop}%
\bibitem [{\citenamefont {Browaeys}\ and\ \citenamefont
  {Lahaye}(2020)}]{Browaeys2020}%
  \BibitemOpen
  \bibfield  {author} {\bibinfo {author} {\bibfnamefont {A.}~\bibnamefont
  {Browaeys}}\ and\ \bibinfo {author} {\bibfnamefont {T.}~\bibnamefont
  {Lahaye}},\ }\bibfield  {title} {\bibinfo {title} {Many-body physics with
  individually controlled {Rydberg} atoms},\ }\href
  {https://doi.org/10.1038/s41567-019-0733-z} {\bibfield  {journal} {\bibinfo
  {journal} {Nature Physics}\ }\textbf {\bibinfo {volume} {16}},\ \bibinfo
  {pages} {132} (\bibinfo {year} {2020})}\BibitemShut {NoStop}%
\bibitem [{\citenamefont {Dutta}\ \emph {et~al.}(2015)\citenamefont {Dutta},
  \citenamefont {Gajda}, \citenamefont {Hauke}, \citenamefont {Lewenstein},
  \citenamefont {Luehmann}, \citenamefont {Malomed}, \citenamefont
  {Sowi\'{n}ski},\ and\ \citenamefont {Zakrzewski}}]{Dutta15}%
  \BibitemOpen
  \bibfield  {author} {\bibinfo {author} {\bibfnamefont {O.}~\bibnamefont
  {Dutta}}, \bibinfo {author} {\bibfnamefont {M.}~\bibnamefont {Gajda}},
  \bibinfo {author} {\bibfnamefont {P.}~\bibnamefont {Hauke}}, \bibinfo
  {author} {\bibfnamefont {M.}~\bibnamefont {Lewenstein}}, \bibinfo {author}
  {\bibfnamefont {D.-S.}\ \bibnamefont {Luehmann}}, \bibinfo {author}
  {\bibfnamefont {B.~A.}\ \bibnamefont {Malomed}}, \bibinfo {author}
  {\bibfnamefont {T.}~\bibnamefont {Sowi\'{n}ski}},\ and\ \bibinfo {author}
  {\bibfnamefont {J.}~\bibnamefont {Zakrzewski}},\ }\bibfield  {title}
  {\bibinfo {title} {Non-standard {H}ubbard models in optical lattices: a
  review},\ }\href {http://stacks.iop.org/0034-4885/78/i=6/a=066001} {\bibfield
   {journal} {\bibinfo  {journal} {Rep. Prog. Phys.}\ }\textbf {\bibinfo
  {volume} {78}},\ \bibinfo {pages} {066001} (\bibinfo {year}
  {2015})}\BibitemShut {NoStop}%
\bibitem [{\citenamefont {Baier}\ \emph {et~al.}(2016)\citenamefont {Baier},
  \citenamefont {Mark}, \citenamefont {Petter}, \citenamefont {Aikawa},
  \citenamefont {Chomaz}, \citenamefont {Cai}, \citenamefont {Baranov},
  \citenamefont {Zoller},\ and\ \citenamefont {Ferlaino}}]{Baier2016}%
  \BibitemOpen
  \bibfield  {author} {\bibinfo {author} {\bibfnamefont {S.}~\bibnamefont
  {Baier}}, \bibinfo {author} {\bibfnamefont {M.~J.}\ \bibnamefont {Mark}},
  \bibinfo {author} {\bibfnamefont {D.}~\bibnamefont {Petter}}, \bibinfo
  {author} {\bibfnamefont {K.}~\bibnamefont {Aikawa}}, \bibinfo {author}
  {\bibfnamefont {L.}~\bibnamefont {Chomaz}}, \bibinfo {author} {\bibfnamefont
  {Z.}~\bibnamefont {Cai}}, \bibinfo {author} {\bibfnamefont {M.}~\bibnamefont
  {Baranov}}, \bibinfo {author} {\bibfnamefont {P.}~\bibnamefont {Zoller}},\
  and\ \bibinfo {author} {\bibfnamefont {F.}~\bibnamefont {Ferlaino}},\
  }\bibfield  {title} {\bibinfo {title} {{Extended Bose-Hubbard models with
  ultracold magnetic atoms}},\ }\href {https://doi.org/10.1126/science.aac9812}
  {\bibfield  {journal} {\bibinfo  {journal} {Science}\ }\textbf {\bibinfo
  {volume} {352}},\ \bibinfo {pages} {201} (\bibinfo {year} {2016})},\ \Eprint
  {https://arxiv.org/abs/https://www.science.org/doi/pdf/10.1126/science.aac9812}
  {https://www.science.org/doi/pdf/10.1126/science.aac9812} \BibitemShut
  {NoStop}%
\bibitem [{\citenamefont {Su}\ \emph {et~al.}(2023)\citenamefont {Su},
  \citenamefont {Douglas}, \citenamefont {Szurek}, \citenamefont {Groth},
  \citenamefont {Ozturk}, \citenamefont {Krahn}, \citenamefont {Hébert},
  \citenamefont {Phelps}, \citenamefont {Ebadi}, \citenamefont {Dickerson},
  \citenamefont {Ferlaino}, \citenamefont {Marković},\ and\ \citenamefont
  {Greiner}}]{Su2023}%
  \BibitemOpen
  \bibfield  {author} {\bibinfo {author} {\bibfnamefont {L.}~\bibnamefont
  {Su}}, \bibinfo {author} {\bibfnamefont {A.}~\bibnamefont {Douglas}},
  \bibinfo {author} {\bibfnamefont {M.}~\bibnamefont {Szurek}}, \bibinfo
  {author} {\bibfnamefont {R.}~\bibnamefont {Groth}}, \bibinfo {author}
  {\bibfnamefont {S.~F.}\ \bibnamefont {Ozturk}}, \bibinfo {author}
  {\bibfnamefont {A.}~\bibnamefont {Krahn}}, \bibinfo {author} {\bibfnamefont
  {A.~H.}\ \bibnamefont {Hébert}}, \bibinfo {author} {\bibfnamefont {G.~A.}\
  \bibnamefont {Phelps}}, \bibinfo {author} {\bibfnamefont {S.}~\bibnamefont
  {Ebadi}}, \bibinfo {author} {\bibfnamefont {S.}~\bibnamefont {Dickerson}},
  \bibinfo {author} {\bibfnamefont {F.}~\bibnamefont {Ferlaino}}, \bibinfo
  {author} {\bibfnamefont {O.}~\bibnamefont {Marković}},\ and\ \bibinfo
  {author} {\bibfnamefont {M.}~\bibnamefont {Greiner}},\ }\href@noop {}
  {\bibinfo {title} {Dipolar quantum solids emerging in a {Hubbard} quantum
  simulator}} (\bibinfo {year} {2023}),\ \Eprint
  {https://arxiv.org/abs/2306.00888} {arXiv:2306.00888 [cond-mat.quant-gas]}
  \BibitemShut {NoStop}%
\bibitem [{\citenamefont {Dalla~Torre}\ \emph {et~al.}(2006)\citenamefont
  {Dalla~Torre}, \citenamefont {Berg},\ and\ \citenamefont
  {Altman}}]{DallaTorre2006}%
  \BibitemOpen
  \bibfield  {author} {\bibinfo {author} {\bibfnamefont {E.~G.}\ \bibnamefont
  {Dalla~Torre}}, \bibinfo {author} {\bibfnamefont {E.}~\bibnamefont {Berg}},\
  and\ \bibinfo {author} {\bibfnamefont {E.}~\bibnamefont {Altman}},\
  }\bibfield  {title} {\bibinfo {title} {Hidden order in 1d {B}ose
  insulators},\ }\href {https://doi.org/10.1103/PhysRevLett.97.260401}
  {\bibfield  {journal} {\bibinfo  {journal} {Phys. Rev. Lett.}\ }\textbf
  {\bibinfo {volume} {97}},\ \bibinfo {pages} {260401} (\bibinfo {year}
  {2006})}\BibitemShut {NoStop}%
\bibitem [{\citenamefont {Rossini}\ and\ \citenamefont
  {Fazio}(2012)}]{Rossini2012}%
  \BibitemOpen
  \bibfield  {author} {\bibinfo {author} {\bibfnamefont {D.}~\bibnamefont
  {Rossini}}\ and\ \bibinfo {author} {\bibfnamefont {R.}~\bibnamefont
  {Fazio}},\ }\bibfield  {title} {\bibinfo {title} {{Phase diagram of the
  extended Bose-Hubbard model}},\ }\href
  {https://doi.org/10.1088/1367-2630/14/6/065012} {\bibfield  {journal}
  {\bibinfo  {journal} {New Journal of Physics}\ }\textbf {\bibinfo {volume}
  {14}},\ \bibinfo {pages} {065012} (\bibinfo {year} {2012})}\BibitemShut
  {NoStop}%
\bibitem [{\citenamefont {Ejima}\ \emph {et~al.}(2014)\citenamefont {Ejima},
  \citenamefont {Lange},\ and\ \citenamefont {Fehske}}]{Ejima2014}%
  \BibitemOpen
  \bibfield  {author} {\bibinfo {author} {\bibfnamefont {S.}~\bibnamefont
  {Ejima}}, \bibinfo {author} {\bibfnamefont {F.}~\bibnamefont {Lange}},\ and\
  \bibinfo {author} {\bibfnamefont {H.}~\bibnamefont {Fehske}},\ }\bibfield
  {title} {\bibinfo {title} {Spectral and entanglement properties of the
  bosonic {Haldane} insulator},\ }\href
  {https://doi.org/10.1103/PhysRevLett.113.020401} {\bibfield  {journal}
  {\bibinfo  {journal} {Phys. Rev. Lett.}\ }\textbf {\bibinfo {volume} {113}},\
  \bibinfo {pages} {020401} (\bibinfo {year} {2014})}\BibitemShut {NoStop}%
\bibitem [{\citenamefont {Pollmann}\ \emph {et~al.}(2010)\citenamefont
  {Pollmann}, \citenamefont {Turner}, \citenamefont {Berg},\ and\ \citenamefont
  {Oshikawa}}]{Pollmann2010}%
  \BibitemOpen
  \bibfield  {author} {\bibinfo {author} {\bibfnamefont {F.}~\bibnamefont
  {Pollmann}}, \bibinfo {author} {\bibfnamefont {A.~M.}\ \bibnamefont
  {Turner}}, \bibinfo {author} {\bibfnamefont {E.}~\bibnamefont {Berg}},\ and\
  \bibinfo {author} {\bibfnamefont {M.}~\bibnamefont {Oshikawa}},\ }\bibfield
  {title} {\bibinfo {title} {Entanglement spectrum of a topological phase in
  one dimension},\ }\href {https://doi.org/10.1103/PhysRevB.81.064439}
  {\bibfield  {journal} {\bibinfo  {journal} {Phys. Rev. B}\ }\textbf {\bibinfo
  {volume} {81}},\ \bibinfo {pages} {064439} (\bibinfo {year}
  {2010})}\BibitemShut {NoStop}%
\bibitem [{\citenamefont {den Nijs}\ and\ \citenamefont
  {Rommelse}(1989)}]{Nijs1989}%
  \BibitemOpen
  \bibfield  {author} {\bibinfo {author} {\bibfnamefont {M.}~\bibnamefont {den
  Nijs}}\ and\ \bibinfo {author} {\bibfnamefont {K.}~\bibnamefont {Rommelse}},\
  }\bibfield  {title} {\bibinfo {title} {Preroughening transitions in crystal
  surfaces and valence-bond phases in quantum spin chains},\ }\href
  {https://doi.org/10.1103/PhysRevB.40.4709} {\bibfield  {journal} {\bibinfo
  {journal} {Phys. Rev. B}\ }\textbf {\bibinfo {volume} {40}},\ \bibinfo
  {pages} {4709} (\bibinfo {year} {1989})}\BibitemShut {NoStop}%
\bibitem [{\citenamefont {Kennedy}\ and\ \citenamefont
  {Tasaki}(1992)}]{Kennedy1992}%
  \BibitemOpen
  \bibfield  {author} {\bibinfo {author} {\bibfnamefont {T.}~\bibnamefont
  {Kennedy}}\ and\ \bibinfo {author} {\bibfnamefont {H.}~\bibnamefont
  {Tasaki}},\ }\bibfield  {title} {\bibinfo {title} {{Hidden
  ${\mathrm{Z}}_{2}$\ifmmode\times\else\texttimes\fi{}${\mathrm{Z}}_{2}$
  symmetry breaking in Haldane-gap antiferromagnets}},\ }\href
  {https://doi.org/10.1103/PhysRevB.45.304} {\bibfield  {journal} {\bibinfo
  {journal} {Phys. Rev. B}\ }\textbf {\bibinfo {volume} {45}},\ \bibinfo
  {pages} {304} (\bibinfo {year} {1992})}\BibitemShut {NoStop}%
\bibitem [{\citenamefont {Fraxanet}\ \emph {et~al.}(2022)\citenamefont
  {Fraxanet}, \citenamefont {Gonz\'alez-Cuadra}, \citenamefont {Pfau},
  \citenamefont {Lewenstein}, \citenamefont {Langen},\ and\ \citenamefont
  {Barbiero}}]{Fraxanet22}%
  \BibitemOpen
  \bibfield  {author} {\bibinfo {author} {\bibfnamefont {J.}~\bibnamefont
  {Fraxanet}}, \bibinfo {author} {\bibfnamefont {D.}~\bibnamefont
  {Gonz\'alez-Cuadra}}, \bibinfo {author} {\bibfnamefont {T.}~\bibnamefont
  {Pfau}}, \bibinfo {author} {\bibfnamefont {M.}~\bibnamefont {Lewenstein}},
  \bibinfo {author} {\bibfnamefont {T.}~\bibnamefont {Langen}},\ and\ \bibinfo
  {author} {\bibfnamefont {L.}~\bibnamefont {Barbiero}},\ }\bibfield  {title}
  {\bibinfo {title} {{Topological Quantum Critical Points in the Extended
  Bose-Hubbard Model}},\ }\href
  {https://doi.org/10.1103/PhysRevLett.128.043402} {\bibfield  {journal}
  {\bibinfo  {journal} {Phys. Rev. Lett.}\ }\textbf {\bibinfo {volume} {128}},\
  \bibinfo {pages} {043402} (\bibinfo {year} {2022})}\BibitemShut {NoStop}%
\bibitem [{\citenamefont {Biedro\'{n}}\ \emph {et~al.}(2018)\citenamefont
  {Biedro\'{n}}, \citenamefont {\L{}\c{a}cki},\ and\ \citenamefont
  {Zakrzewski}}]{Biedron18}%
  \BibitemOpen
  \bibfield  {author} {\bibinfo {author} {\bibfnamefont {K.}~\bibnamefont
  {Biedro\'{n}}}, \bibinfo {author} {\bibfnamefont {M.}~\bibnamefont
  {\L{}\c{a}cki}},\ and\ \bibinfo {author} {\bibfnamefont {J.}~\bibnamefont
  {Zakrzewski}},\ }\bibfield  {title} {\bibinfo {title} {Extended
  {B}ose-{H}ubbard model with dipolar and contact interactions},\ }\href
  {https://doi.org/10.1103/PhysRevB.97.245102} {\bibfield  {journal} {\bibinfo
  {journal} {Phys. Rev. B}\ }\textbf {\bibinfo {volume} {97}},\ \bibinfo
  {pages} {245102} (\bibinfo {year} {2018})}\BibitemShut {NoStop}%
\bibitem [{\citenamefont {Kraus}\ \emph {et~al.}(2020)\citenamefont {Kraus},
  \citenamefont {Biedro\ifmmode~\acute{n}\else \'{n}\fi{}}, \citenamefont
  {Zakrzewski},\ and\ \citenamefont {Morigi}}]{Kraus20}%
  \BibitemOpen
  \bibfield  {author} {\bibinfo {author} {\bibfnamefont {R.}~\bibnamefont
  {Kraus}}, \bibinfo {author} {\bibfnamefont {K.}~\bibnamefont
  {Biedro\ifmmode~\acute{n}\else \'{n}\fi{}}}, \bibinfo {author} {\bibfnamefont
  {J.}~\bibnamefont {Zakrzewski}},\ and\ \bibinfo {author} {\bibfnamefont
  {G.}~\bibnamefont {Morigi}},\ }\bibfield  {title} {\bibinfo {title}
  {Superfluid phases induced by dipolar interactions},\ }\href
  {https://doi.org/10.1103/PhysRevB.101.174505} {\bibfield  {journal} {\bibinfo
   {journal} {Phys. Rev. B}\ }\textbf {\bibinfo {volume} {101}},\ \bibinfo
  {pages} {174505} (\bibinfo {year} {2020})}\BibitemShut {NoStop}%
\bibitem [{\citenamefont {Wall}\ and\ \citenamefont {Carr}(2013)}]{Wall2013}%
  \BibitemOpen
  \bibfield  {author} {\bibinfo {author} {\bibfnamefont {M.~L.}\ \bibnamefont
  {Wall}}\ and\ \bibinfo {author} {\bibfnamefont {L.~D.}\ \bibnamefont
  {Carr}},\ }\bibfield  {title} {\bibinfo {title} {Dipole{\textendash}dipole
  interactions in optical lattices do not follow an inverse cube power law},\
  }\href {https://doi.org/10.1088/1367-2630/15/12/123005} {\bibfield  {journal}
  {\bibinfo  {journal} {New Journal of Physics}\ }\textbf {\bibinfo {volume}
  {15}},\ \bibinfo {pages} {123005} (\bibinfo {year} {2013})}\BibitemShut
  {NoStop}%
\bibitem [{\citenamefont {Korbmacher}\ \emph
  {et~al.}(2023{\natexlab{a}})\citenamefont {Korbmacher}, \citenamefont
  {Sierant}, \citenamefont {Li}, \citenamefont {Deng}, \citenamefont
  {Zakrzewski},\ and\ \citenamefont {Santos}}]{Korbmacher2023}%
  \BibitemOpen
  \bibfield  {author} {\bibinfo {author} {\bibfnamefont {H.}~\bibnamefont
  {Korbmacher}}, \bibinfo {author} {\bibfnamefont {P.}~\bibnamefont {Sierant}},
  \bibinfo {author} {\bibfnamefont {W.}~\bibnamefont {Li}}, \bibinfo {author}
  {\bibfnamefont {X.}~\bibnamefont {Deng}}, \bibinfo {author} {\bibfnamefont
  {J.}~\bibnamefont {Zakrzewski}},\ and\ \bibinfo {author} {\bibfnamefont
  {L.}~\bibnamefont {Santos}},\ }\bibfield  {title} {\bibinfo {title} {Lattice
  control of nonergodicity in a polar lattice gas},\ }\href
  {https://doi.org/10.1103/PhysRevA.107.013301} {\bibfield  {journal} {\bibinfo
   {journal} {Phys. Rev. A}\ }\textbf {\bibinfo {volume} {107}},\ \bibinfo
  {pages} {013301} (\bibinfo {year} {2023}{\natexlab{a}})}\BibitemShut
  {NoStop}%
\bibitem [{\citenamefont {Korbmacher}\ \emph
  {et~al.}(2023{\natexlab{b}})\citenamefont {Korbmacher}, \citenamefont
  {Dom\'{\i}nguez-Castro}, \citenamefont {Li}, \citenamefont {Zakrzewski},\
  and\ \citenamefont {Santos}}]{PhysRevA.107.063307}%
  \BibitemOpen
  \bibfield  {author} {\bibinfo {author} {\bibfnamefont {H.}~\bibnamefont
  {Korbmacher}}, \bibinfo {author} {\bibfnamefont {G.~A.}\ \bibnamefont
  {Dom\'{\i}nguez-Castro}}, \bibinfo {author} {\bibfnamefont {W.-H.}\
  \bibnamefont {Li}}, \bibinfo {author} {\bibfnamefont {J.}~\bibnamefont
  {Zakrzewski}},\ and\ \bibinfo {author} {\bibfnamefont {L.}~\bibnamefont
  {Santos}},\ }\bibfield  {title} {\bibinfo {title} {Transversal effects on the
  ground state of hard-core dipolar bosons in one-dimensional optical
  lattices},\ }\href {https://doi.org/10.1103/PhysRevA.107.063307} {\bibfield
  {journal} {\bibinfo  {journal} {Phys. Rev. A}\ }\textbf {\bibinfo {volume}
  {107}},\ \bibinfo {pages} {063307} (\bibinfo {year}
  {2023}{\natexlab{b}})}\BibitemShut {NoStop}%
\bibitem [{\citenamefont {Gross}\ and\ \citenamefont {Bakr}(2021)}]{Gross2021}%
  \BibitemOpen
  \bibfield  {author} {\bibinfo {author} {\bibfnamefont {C.}~\bibnamefont
  {Gross}}\ and\ \bibinfo {author} {\bibfnamefont {W.~S.}\ \bibnamefont
  {Bakr}},\ }\bibfield  {title} {\bibinfo {title} {Quantum gas microscopy for
  single atom and spin detection},\ }\href
  {https://doi.org/10.1038/s41567-021-01370-5} {\bibfield  {journal} {\bibinfo
  {journal} {Nature Physics}\ }\textbf {\bibinfo {volume} {17}},\ \bibinfo
  {pages} {1316} (\bibinfo {year} {2021})}\BibitemShut {NoStop}%
\bibitem [{\citenamefont {McCulloch}(2008)}]{Mcculloch08}%
  \BibitemOpen
  \bibfield  {author} {\bibinfo {author} {\bibfnamefont {I.~P.}\ \bibnamefont
  {McCulloch}},\ }\href@noop {} {\bibinfo {title} {Infinite size density matrix
  renormalization group, revisited}} (\bibinfo {year} {2008}),\ \Eprint
  {https://arxiv.org/abs/0804.2509} {arXiv:0804.2509 [cond-mat.str-el]}
  \BibitemShut {NoStop}%
\bibitem [{\citenamefont {Sowi\ifmmode~\acute{n}\else \'{n}\fi{}ski}\ \emph
  {et~al.}(2012)\citenamefont {Sowi\ifmmode~\acute{n}\else \'{n}\fi{}ski},
  \citenamefont {Dutta}, \citenamefont {Hauke}, \citenamefont {Tagliacozzo},\
  and\ \citenamefont {Lewenstein}}]{Sowinski2012}%
  \BibitemOpen
  \bibfield  {author} {\bibinfo {author} {\bibfnamefont {T.}~\bibnamefont
  {Sowi\ifmmode~\acute{n}\else \'{n}\fi{}ski}}, \bibinfo {author}
  {\bibfnamefont {O.}~\bibnamefont {Dutta}}, \bibinfo {author} {\bibfnamefont
  {P.}~\bibnamefont {Hauke}}, \bibinfo {author} {\bibfnamefont
  {L.}~\bibnamefont {Tagliacozzo}},\ and\ \bibinfo {author} {\bibfnamefont
  {M.}~\bibnamefont {Lewenstein}},\ }\bibfield  {title} {\bibinfo {title}
  {Dipolar molecules in optical lattices},\ }\href
  {https://doi.org/10.1103/PhysRevLett.108.115301} {\bibfield  {journal}
  {\bibinfo  {journal} {Phys. Rev. Lett.}\ }\textbf {\bibinfo {volume} {108}},\
  \bibinfo {pages} {115301} (\bibinfo {year} {2012})}\BibitemShut {NoStop}%
\bibitem [{\citenamefont {Kraus}\ \emph {et~al.}(2022)\citenamefont {Kraus},
  \citenamefont {Chanda}, \citenamefont {Zakrzewski},\ and\ \citenamefont
  {Morigi}}]{Kraus22}%
  \BibitemOpen
  \bibfield  {author} {\bibinfo {author} {\bibfnamefont {R.}~\bibnamefont
  {Kraus}}, \bibinfo {author} {\bibfnamefont {T.}~\bibnamefont {Chanda}},
  \bibinfo {author} {\bibfnamefont {J.}~\bibnamefont {Zakrzewski}},\ and\
  \bibinfo {author} {\bibfnamefont {G.}~\bibnamefont {Morigi}},\ }\bibfield
  {title} {\bibinfo {title} {Quantum phases of dipolar bosons in
  one-dimensional optical lattices},\ }\href
  {https://doi.org/10.1103/PhysRevB.106.035144} {\bibfield  {journal} {\bibinfo
   {journal} {Phys. Rev. B}\ }\textbf {\bibinfo {volume} {106}},\ \bibinfo
  {pages} {035144} (\bibinfo {year} {2022})}\BibitemShut {NoStop}%
\bibitem [{\citenamefont {{\L}{\c{a}}cki}\ \emph {et~al.}(2013)\citenamefont
  {{\L}{\c{a}}cki}, \citenamefont {Delande},\ and\ \citenamefont
  {Zakrzewski}}]{Lacki13}%
  \BibitemOpen
  \bibfield  {author} {\bibinfo {author} {\bibfnamefont {M.}~\bibnamefont
  {{\L}{\c{a}}cki}}, \bibinfo {author} {\bibfnamefont {D.}~\bibnamefont
  {Delande}},\ and\ \bibinfo {author} {\bibfnamefont {J.}~\bibnamefont
  {Zakrzewski}},\ }\bibfield  {title} {\bibinfo {title} {Dynamics of cold
  bosons in optical lattices: effects of higher bloch bands},\ }\href
  {https://doi.org/10.1088/1367-2630/15/1/013062} {\bibfield  {journal}
  {\bibinfo  {journal} {New Journal of Physics}\ }\textbf {\bibinfo {volume}
  {15}},\ \bibinfo {pages} {013062} (\bibinfo {year} {2013})}\BibitemShut
  {NoStop}%
\bibitem [{\citenamefont {Hughes}\ \emph {et~al.}(2023)\citenamefont {Hughes},
  \citenamefont {Lode}, \citenamefont {Jaksch},\ and\ \citenamefont
  {Molignini}}]{Hughes23}%
  \BibitemOpen
  \bibfield  {author} {\bibinfo {author} {\bibfnamefont {M.}~\bibnamefont
  {Hughes}}, \bibinfo {author} {\bibfnamefont {A.~U.~J.}\ \bibnamefont {Lode}},
  \bibinfo {author} {\bibfnamefont {D.}~\bibnamefont {Jaksch}},\ and\ \bibinfo
  {author} {\bibfnamefont {P.}~\bibnamefont {Molignini}},\ }\bibfield  {title}
  {\bibinfo {title} {Accuracy of quantum simulators with ultracold dipolar
  molecules: A quantitative comparison between continuum and lattice
  descriptions},\ }\href {https://doi.org/10.1103/PhysRevA.107.033323}
  {\bibfield  {journal} {\bibinfo  {journal} {Phys. Rev. A}\ }\textbf {\bibinfo
  {volume} {107}},\ \bibinfo {pages} {033323} (\bibinfo {year}
  {2023})}\BibitemShut {NoStop}%
\bibitem [{\citenamefont {Batrouni}\ \emph {et~al.}(2014)\citenamefont
  {Batrouni}, \citenamefont {Rousseau}, \citenamefont {Scalettar},\ and\
  \citenamefont {Gr\'emaud}}]{Batrouni2014}%
  \BibitemOpen
  \bibfield  {author} {\bibinfo {author} {\bibfnamefont {G.~G.}\ \bibnamefont
  {Batrouni}}, \bibinfo {author} {\bibfnamefont {V.~G.}\ \bibnamefont
  {Rousseau}}, \bibinfo {author} {\bibfnamefont {R.~T.}\ \bibnamefont
  {Scalettar}},\ and\ \bibinfo {author} {\bibfnamefont {B.}~\bibnamefont
  {Gr\'emaud}},\ }\bibfield  {title} {\bibinfo {title} {Competing phases, phase
  separation, and coexistence in the extended one-dimensional bosonic {Hubbard}
  model},\ }\href {https://doi.org/10.1103/PhysRevB.90.205123} {\bibfield
  {journal} {\bibinfo  {journal} {Phys. Rev. B}\ }\textbf {\bibinfo {volume}
  {90}},\ \bibinfo {pages} {205123} (\bibinfo {year} {2014})}\BibitemShut
  {NoStop}%
\bibitem [{\citenamefont {Kottmann}\ \emph {et~al.}(2021)\citenamefont
  {Kottmann}, \citenamefont {Haller}, \citenamefont {Ac\'{\i}n}, \citenamefont
  {Astrakharchik},\ and\ \citenamefont {Lewenstein}}]{Kottmann2021}%
  \BibitemOpen
  \bibfield  {author} {\bibinfo {author} {\bibfnamefont {K.}~\bibnamefont
  {Kottmann}}, \bibinfo {author} {\bibfnamefont {A.}~\bibnamefont {Haller}},
  \bibinfo {author} {\bibfnamefont {A.}~\bibnamefont {Ac\'{\i}n}}, \bibinfo
  {author} {\bibfnamefont {G.~E.}\ \bibnamefont {Astrakharchik}},\ and\
  \bibinfo {author} {\bibfnamefont {M.}~\bibnamefont {Lewenstein}},\ }\bibfield
   {title} {\bibinfo {title} {{Supersolid-superfluid phase separation in the
  extended Bose-Hubbard model}},\ }\href
  {https://doi.org/10.1103/PhysRevB.104.174514} {\bibfield  {journal} {\bibinfo
   {journal} {Phys. Rev. B}\ }\textbf {\bibinfo {volume} {104}},\ \bibinfo
  {pages} {174514} (\bibinfo {year} {2021})}\BibitemShut {NoStop}%
\bibitem [{\citenamefont {Chen}\ \emph {et~al.}(2003)\citenamefont {Chen},
  \citenamefont {Hida},\ and\ \citenamefont {Sanctuary}}]{Chen2003}%
  \BibitemOpen
  \bibfield  {author} {\bibinfo {author} {\bibfnamefont {W.}~\bibnamefont
  {Chen}}, \bibinfo {author} {\bibfnamefont {K.}~\bibnamefont {Hida}},\ and\
  \bibinfo {author} {\bibfnamefont {B.~C.}\ \bibnamefont {Sanctuary}},\
  }\bibfield  {title} {\bibinfo {title} {Ground-state phase diagram of $s=1$
  $\mathrm{XXZ}$ chains with uniaxial single-ion-type anisotropy},\ }\href
  {https://doi.org/10.1103/PhysRevB.67.104401} {\bibfield  {journal} {\bibinfo
  {journal} {Phys. Rev. B}\ }\textbf {\bibinfo {volume} {67}},\ \bibinfo
  {pages} {104401} (\bibinfo {year} {2003})}\BibitemShut {NoStop}%
\bibitem [{foo()}]{footnote-R}%
  \BibitemOpen
  \href@noop {} {}\bibinfo {note} {Additional precaution is necessary due to
  the oscillatory character of the string-like correlations in the $\ell$DW
  phases. Hence, before taking the limit $j\rightarrow\infty$ we first average
  over few $j$ sites respecting the phase periodicity. In this way the
  oscillations in $R(q)$ in $\ell$DW phases cancel out to zero.}\BibitemShut
  {Stop}%
\bibitem [{\citenamefont {Calabrese}\ and\ \citenamefont
  {Lefevre}(2008)}]{Calabrese2008}%
  \BibitemOpen
  \bibfield  {author} {\bibinfo {author} {\bibfnamefont {P.}~\bibnamefont
  {Calabrese}}\ and\ \bibinfo {author} {\bibfnamefont {A.}~\bibnamefont
  {Lefevre}},\ }\bibfield  {title} {\bibinfo {title} {Entanglement spectrum in
  one-dimensional systems},\ }\href
  {https://doi.org/10.1103/PhysRevA.78.032329} {\bibfield  {journal} {\bibinfo
  {journal} {Phys. Rev. A}\ }\textbf {\bibinfo {volume} {78}},\ \bibinfo
  {pages} {032329} (\bibinfo {year} {2008})}\BibitemShut {NoStop}%
\bibitem [{\citenamefont {Berg}\ \emph {et~al.}(2008)\citenamefont {Berg},
  \citenamefont {Dalla~Torre}, \citenamefont {Giamarchi},\ and\ \citenamefont
  {Altman}}]{Berg2008}%
  \BibitemOpen
  \bibfield  {author} {\bibinfo {author} {\bibfnamefont {E.}~\bibnamefont
  {Berg}}, \bibinfo {author} {\bibfnamefont {E.~G.}\ \bibnamefont
  {Dalla~Torre}}, \bibinfo {author} {\bibfnamefont {T.}~\bibnamefont
  {Giamarchi}},\ and\ \bibinfo {author} {\bibfnamefont {E.}~\bibnamefont
  {Altman}},\ }\bibfield  {title} {\bibinfo {title} {Rise and fall of hidden
  string order of lattice bosons},\ }\href
  {https://doi.org/10.1103/PhysRevB.77.245119} {\bibfield  {journal} {\bibinfo
  {journal} {Phys. Rev. B}\ }\textbf {\bibinfo {volume} {77}},\ \bibinfo
  {pages} {245119} (\bibinfo {year} {2008})}\BibitemShut {NoStop}%
\bibitem [{\citenamefont {Dotsenko}(2020)}]{Dotsenko2020}%
  \BibitemOpen
  \bibfield  {author} {\bibinfo {author} {\bibfnamefont {V.~S.}\ \bibnamefont
  {Dotsenko}},\ }\bibfield  {title} {\bibinfo {title} {{Four spins correlation
  function of the q states Potts model, for general values of q. Its
  percolation model limit q$\rightarrow$1}},\ }\href
  {https://doi.org/https://doi.org/10.1016/j.nuclphysb.2020.114973} {\bibfield
  {journal} {\bibinfo  {journal} {Nuclear Physics B}\ }\textbf {\bibinfo
  {volume} {953}},\ \bibinfo {pages} {114973} (\bibinfo {year}
  {2020})}\BibitemShut {NoStop}%
\end{thebibliography}
